\documentclass[journal]{IEEEtran}
\usepackage{cite}

\usepackage{lipsum}
\ifCLASSINFOpdf
  \usepackage[pdftex]{graphicx}
\else
\fi
\usepackage{amsmath}
\begin{document}

\IEEEoverridecommandlockouts
\IEEEpubid{\makebox[\columnwidth]{978-1-5386-5541-2/18/\$31.00~\copyright2018 IEEE \hfill} \hspace{\columnsep}\makebox[\columnwidth]{ }}
\title{Observation of Long Turn-on Delay in Pulsed Quantum Cascade Lasers}
\IEEEpubidadjcol
%

\author{E.~D.~Cherotchenko, V.~V.~Dudelev, D.~A.~Mikhailov, S.~N.~Losev, A.~V.~Babichev, A.~G.~Gladyshev, I.~I.~Novikov, A.~V.~Lutetskiy, D.~A.~Veselov, S.~O.~Slipchenko, N.~A.~Pikhtin, L.~Ya.~Karachinsky, D.~V.~Denisov, V.~I.~Kuchinskii, E.~A.~Kognovitskaya, A.~Yu.~Egorov, R.~Teissier, A.~N.~Baranov, G.~S.~Sokolovskii (IEEE member).
\thanks{
E. D. Cherotchenko, V. V. Dudelev, D. A. Mikhailov, S. N. Losev, A. V. Lutetskiy, D. A. Veselov, S. O. Slipchenko, N. A. Pikhtin, V. I. Kuchinskii, G. S. Sokolovskii are with Ioffe institute, St. Petersburg, Russia.

 E. A. Kognovitskaya is with St.   Petersburg  Electrotechnical  University  “LETI”, St.Petersburg,  Russia and with Ioffe Institute, Russia.
 
 I. I. Novikov, L. Ya. Karachinsky are with  Ioffe institute, ITMO University, Russia and Connector  Optics  LLC, Saint  Petersburg,  Russia.
 
 A. G. Gladyshev and D. V. Denisov are with Connector  Optics  LLC, Saint  Petersburg,  Russia.
 
 A. V. Babichev, is with ITMO University, Russia and Connector  Optics  LLC, Saint  Petersburg,  Russia.
 
 R. Teissier, A. N. Baranov are with the University of Montpellier, France.

-e-mail: echerotchenko@gmail.com}
\thanks{Manuscript received July 12, 2021;}}

\maketitle

\begin{abstract}
We present an experimental study of the turn-on delay in pulsed mid-infrared quantum cascade lasers. We report the unexpectedly long delay time depending on the pumping current, which does not agree with conventional theoretical predictions for step-like excitation. Similar behavior has been observed in InP- and InAs-based QCLs emitting near 8 µm. Numerical simulations performed using a model based on rate equations for excitation by current pulses with non-zero rise time provide fair agreement with our observations.
\end{abstract}


%
\IEEEpeerreviewmaketitle

\section{Introduction}
%
%
%
%
\IEEEPARstart{A}{lmost} three decades after the first practical realization, quantum cascade lasers (QCLs), first proposed by Kazarinov and Suris \cite{kazarinov1971}, evolved from a proof-of-principle sample  \cite{faist1994} to a widespread technology, which is used in a huge variety of applications, such as trace-chemical sensing \cite{fuchs2010}, hyper-spectral imaging\cite{goyal2014}, free-space communications\cite{chuanwei2015, Delga2019}, health monitoring\cite{wojtas2012}, spectroscopy \cite{Schrottke2020} and many others.
Recent investigations in topologically protected systems allowed creation of unique QCLs based on valley edge states\cite{Zeng2020} paving the way to a new type of devices that are ultimately robust to the structural defects. 
At the same time, despite of significant developments in this cutting edge research, some aspects of the QCL physics remain vastly unrevealed.
One of them is the response of a QCL to pulsed current excitation. The turn-on delay time is a key parameter in such applications as LIDARs and telecommunication, where the switching time defines the information transfer rate.

Up to now, the turn-on delay in QCLs was mainly studied theoretically~\cite{hamadou2009, hamadou2013, yong2013, yong2018, dudelev2018} in the framework of rate equations for two-level or three-level systems \cite{agnew2016, kundu2018, choi2008, webb2014, hamadou2018, Ashok2019}.
The main conclusions of these discussions are extremely short turn-on delays of the order of few picoseconds and relatively short build-up times, i.e. the time necessary for the laser to reach 10$\%$ of its stationary photon mode occupancy. The last were estimated to be well below hundred picoseconds at double threshold pumping\cite{hamadou2009}.  At the same time, all above mentioned works assumed step-like or periodic square pump pulses, whereas realistic excitation pulses have non-zero rise times. Moreover, the experimental confirmations of such a short delay times were not made.

In our work we demonstrate the measurements of the turn-on delay in two structurally different sets of mid-infrared QCLs. 
Unexpectedly, we find that the delay times are orders of magnitude longer which is in contradiction with the predictions of the existing models. Apart form that the turn-on delay-current relation demonstrates counterintuitive non-monotonic behavior with maximum appearing well above the threshold. This observation also contradicts with the predicted hyperbola-like shape, with maximum value at the threshold.
The proposed theoretical model of the QCL response to pump pulses with non-zero rise time is in qualitative agreement with the observed dependence of the turn-on delay on the pumping amplitude.

\section{Experimental setup and measurements}
\label{sec:level1}

\begin{figure}[ht]
\centering
\includegraphics[width=1\linewidth]{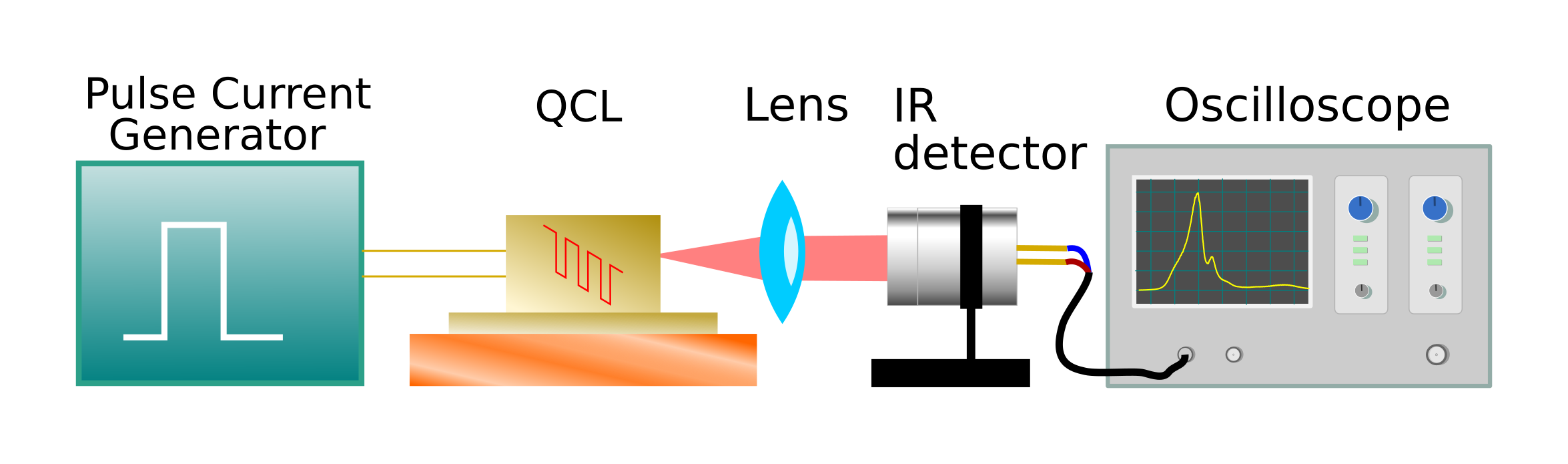}
\caption{Schematic illustration of the experimental setup.}
\label{Scheme}
\end{figure}

The schematic illustration of the experimental setup used for the turn-on delay measurements is shown in Fig.~\ref{Scheme}. It consists of a pulse generator, the Vigo PVI-4TE-10.6 photodetector, a BaF$_2$ lens, and the Infinium Agilent 54854A oscilloscope. The lasers were driven using 150-ns-long current pulses with 20 ns rise time at 25 kHz repetition rate. The current-light signals position accuracy was 10 ps, while the photodetector and oscilloscope cut-off frequency was 1 GHz that translates to overall resolution of about 1 ns. The current was measured through a series resistance. The laser emission was focused onto the sensor, directly connected to the oscilloscope. The optical and electrical path lengths were carefully measured, and the difference between the two lengths was taken into account. Spectral measurements were carried out with a MDR23 monochromator (not shown in the scheme).

\begin{figure}[b]
\centering
\includegraphics[width=0.7\linewidth]{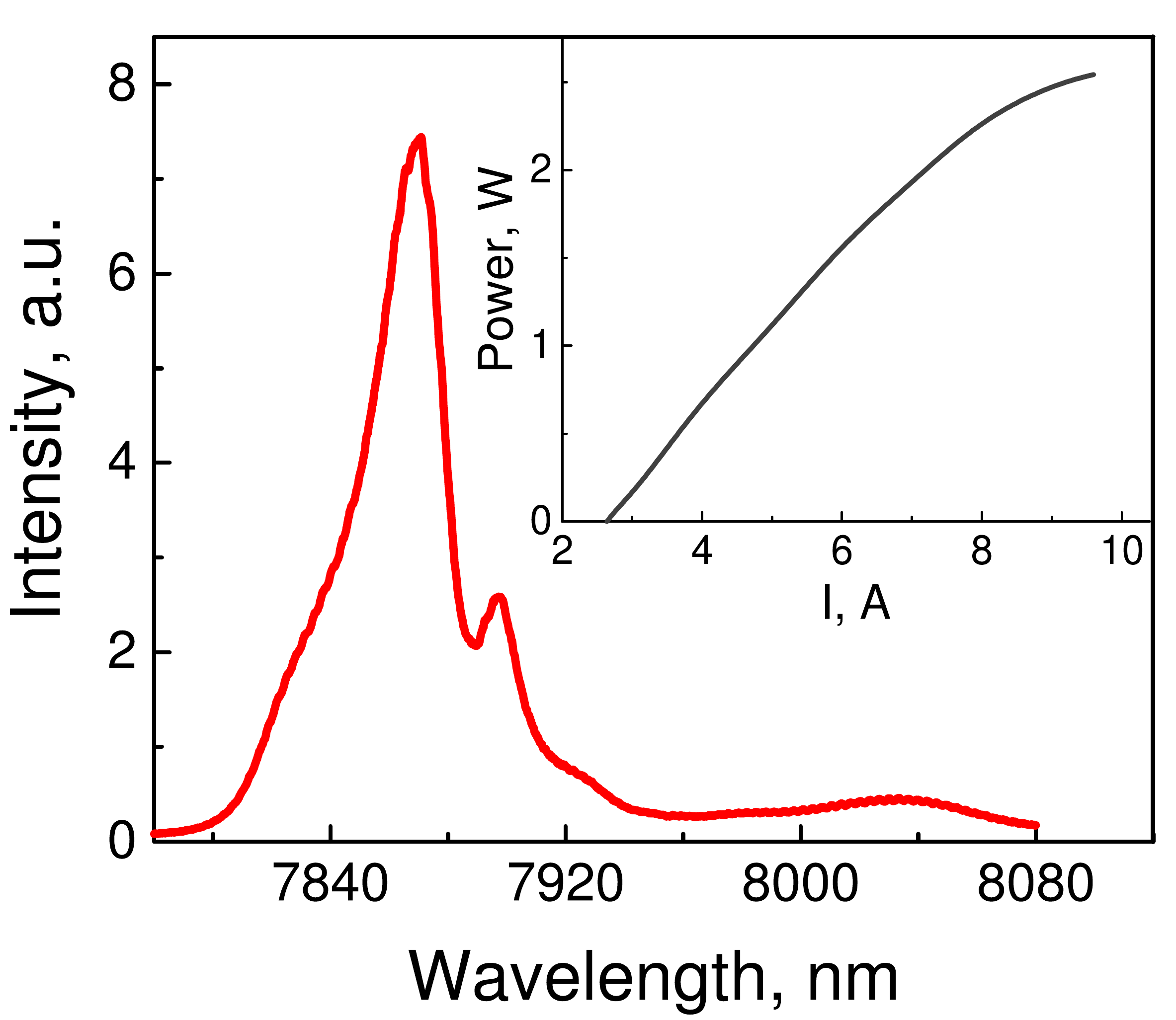}
\caption{(color online) Spectrum and light-current (inset) characteristics of the QCL sample, corresponding to Fig.\ref{TS5020}a. The sample has a $3\,$mm length, 20$\,\mu$m stripe width, and 50 cascades in the active layer.}
\label{Spectrum}
\end{figure}
\begin{figure}[t]
\centering
\includegraphics[width=0.9\linewidth]{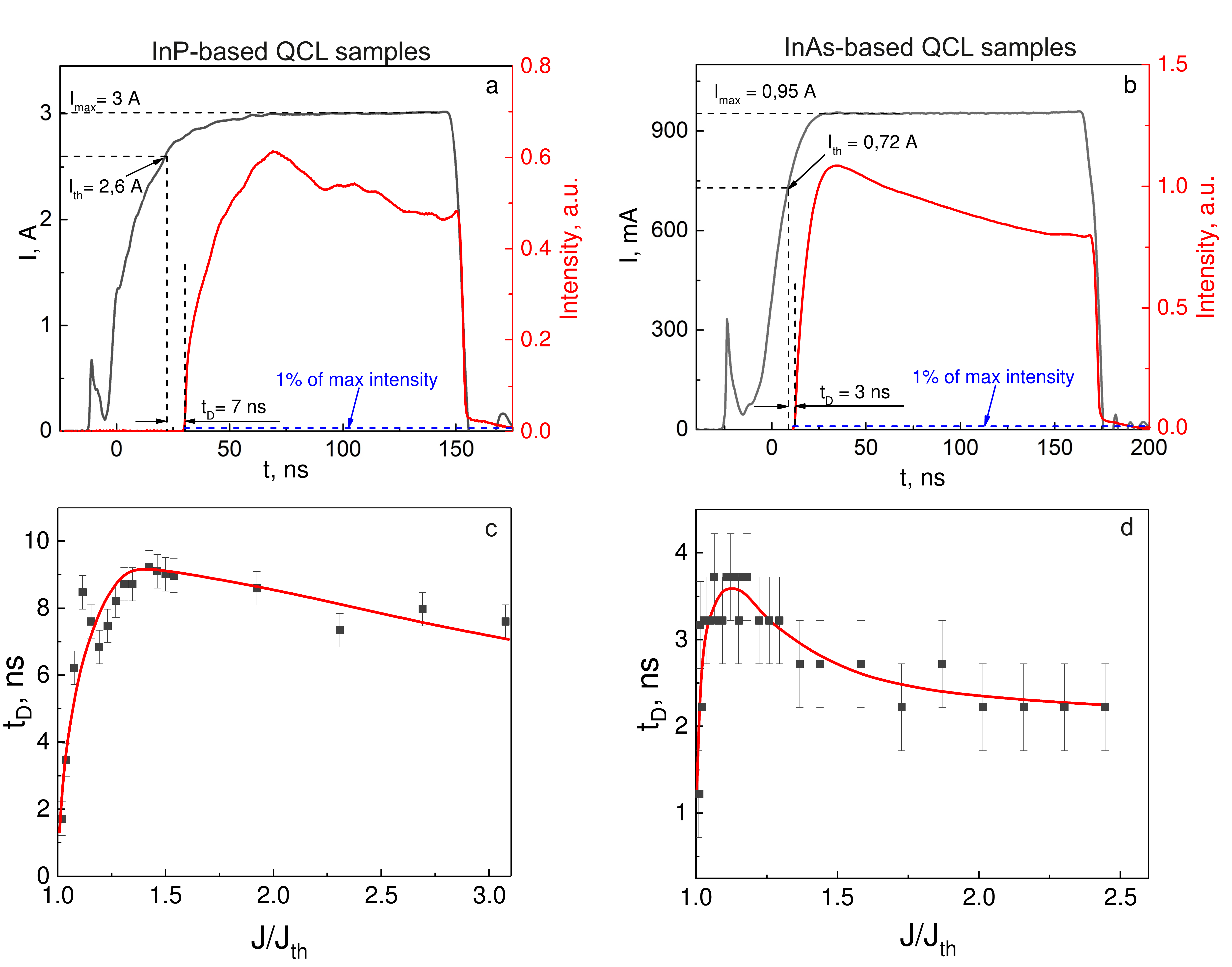}
\caption{(color online) a) Shape of the pumping current (black) and generated light (red) pulses of the InP QCL samples. Dashed lines show maximum current amplitude, threshold current (black horizontal lines), 1\% of the maximum lasing intensity (blue dashed line) and the turn-on delay; b) Same as (a) for InAs QCL.  c) Delay time as a function of current normalized to threshold, measured for InP QCL samples. Solid line is given as a guide to the eye; d) Same as (c) for InAs QCL. }
\label{TS5020}
\end{figure}

The first experiments were performed on lattice-matched InP-based QCLs similar to those reported in Refs.\cite{babichev2019} and \cite{dudelev2020}.
The QCLs grown by molecular beam epitaxy were based on the diagonal two-phonon resonant design with 50 cascades in the active region\cite{babichev2017}.
At room temperature, they emitted at 7.9 $\mu$m with a threshold current density of 3-5 kA/cm$^2$ and maximum peak output power exceeding 1 W per facet\cite{babichev2019, dudelev2020}.
The studied devices with a ridge width of 20 and 50 $\mu$m were 3 mm long.
Typical low-resolution emission spectrum and the light-current characteristic of the lasers are shown in Fig.~\ref{Spectrum}.
Fig.~\ref{TS5020}a shows the measured shape of the current and light pulses. 
The horizontal dashed lines indicate the maximum current amplitude for the given measurement and the threshold current. The latter was defined as the intercept of the tangent line with the horizontal axis on light-current characteristic, of typical shape presented in fig.~\ref{Spectrum}. The threshold was reached during the current rising, before the plateau is achieved.

In  these  experiments  the  turn-on  delay  was  defined  as  the  time  from pumping current reaching  the  laser  threshold  to  the  moment,  when  the  lasing  intensity reaches  1\%  of  its maximum  (shown  by  the  vertical  black  dashed  lines  and  horizontal  blue  dashed  line on the  graph). 
The results are demonstrated in Fig.\ref{TS5020}c.
The measured turn-on delay time is surprisingly long reaching almost 10 ns, which is few orders of magnitude longer comparing to the theoretical predictions for step-like pumping \cite{hamadou2009,hamadou2013,yong2013,yong2018,dudelev2018,agnew2016,kundu2018,choi2008,webb2014,hamadou2018,Ashok2019}.
Moreover, the dependence of the turn-on delay on the pump current exhibits a non-monotonic behavior.
It sharply rises at low currents and reaches the maximum at J/J$_{th}$ values of about 1.5 and then slowly decreases with further increase in pumping current.
This behavior is also in contradiction with the theoretical models predicting the maximum time delay at the threshold current~\cite{hamadou2009, hamadou2013, yong2013, yong2018}.

To exclude artificial technology-related factors that could affect the dynamic characteristics of the studied devices and be responsible for the observed deviations from the expected behavior we performed the same measurements on completely different QCLs based on other materials and fabricated in other facilities. 
These were InAs/AlSb QCLs similar to those reported in Ref.\cite{baranov2016}.

The InAs QCL design with vertical transitions was modified to shift the emission to 8 $\mu$m like in the studied InP-based devices.
The structure with 40 cascade stages in the active region was grown by molecular beam epitaxy and processed to 3.6-mm-long and 17-$\mu$m-wide ridge lasers.
Hard baked photoresist was used for electrical insulation in these devices, in contrast to SiO$_2$ in the InP-based QCLs discussed earlier.
The lasers emitted at 8.2 $\mu$m with a threshold current of 0.7-0.8 A (J$_{th}$ $\approx$ 1.3 kA/cm$^2$) in pulsed mode at room temperature. 
Results obtained with the InAs-based QCL are shown in Fig.~\ref{TS5020}b and \ref{TS5020}d.
The measured characteristics are very similar to those found for the InP-based lasers discussed above despite the enormous difference in the design and technology of the studied devices.
Similarly to the case of InP QCLs, the measured value of the turn-on delay of 4 ns is longer than the value expected from the theoretical predictions and its non-monotonous dependence on the pumping current is in qualitative contradiction with the earlier model predictions.

\section{Numerical simulations}
\label{sec:coupled}
\begin{table}
\caption{\label{tab:table1} Parameters used in the simulations. Indices "3", "2", and "1" indicate the upper and lower levels and ground state correspondingly. Double indices indicate the initial and final states.}


\begin{tabular}{|c|c|c|}
 Parameter&Meaning&Value\\ \hline
 J&Current density&-\\
 W&Lateral width of the cavity
 &20$\mu$m\\
 $K$&Number of cascades
 & 50\\
 $\tau_{sp}$&Spontaneous relaxation time
 &38 ns\\
  $\tau_{3}$&Lifetime on the upper level
 &1.4 ps\\
  $\tau_{31}$&Scattering time due to LO-phonon emission
 &4.2 ps\\
  $\tau_{21}$&Scattering time due to LO-phonon emission
 &0.3 ps\\
  $\tau_{32}$&Scattering time due to LO-phonon emission
 &2.1 ps\\
  $\tau_{p}$&Photon lifetime
 &3.36 ps\\
  $\tau_{out}$&Electron escape time
 &1 ps\\
  $\sigma_{32}$&Stimulated emission crossection &1.85$\times10^{-18}$\\
  $\beta$&Spontaneous emission factor
 &2$\times10^{-3}$\\
  $\Gamma$&Mode confinement factor
 &0.32\\
  $V$&Volume of the active area
 &-\\
  $c'$&Speed of Light in the medium
 &0.3$c$\\
 \end{tabular}
\end{table}

Our model used in simulations is based on the system of rate equations for the  three-level
system under electrical injection \cite{hamadou2009}. We studied the QCL response to
pulse excitation when the current smoothly increased from zero to its
maximum value  (in contrast to a piecewise constant function typically
used for mathematical description of a step-like excitation):
\begin{subequations}
\label{RateEq}
\begin{equation}
\frac{dN_3}{dt} = WL\frac{J}{e}-\frac{N_3}{\tau_3}-S,\label{subeq:1}
\end{equation}
\begin{eqnarray}
\frac{dN_2}{dt} = \left(\frac{N_3}{\tau_{32}}+\frac{N_3}{\tau_{sp}}\right)-\frac{N_2}{\tau_{21}}+S,\label{subeq:2}
\end{eqnarray}
\begin{equation}
\frac{dN_1}{dt} = \frac{N_3}{\tau_{31}}+\frac{N_2}{\tau_{21}}-\frac{N_1}{\tau_{out}},\label{subeq:3}
\end{equation}
\begin{equation}
\frac{dN_{ph}}{dt} =KS+K\beta\frac{N_3}{\tau_{sp}}-\frac{N_{ph}}{\tau_p},\label{subeq:4}
\end{equation}
\end{subequations}

Here $N_3$ and $N_2$ denote the occupation numbers of the upper and lower excited states respectively, $N_1$ stands for the ground state occupation, and $N_{ph}$ is the occupation number of the photonic mode, $ S=\Gamma\frac{c'\sigma_{32}}{V}(N_3-N_2)N_{ph}$.
\begin{figure}[t]
    \centering
    \includegraphics[width=0.9\linewidth]{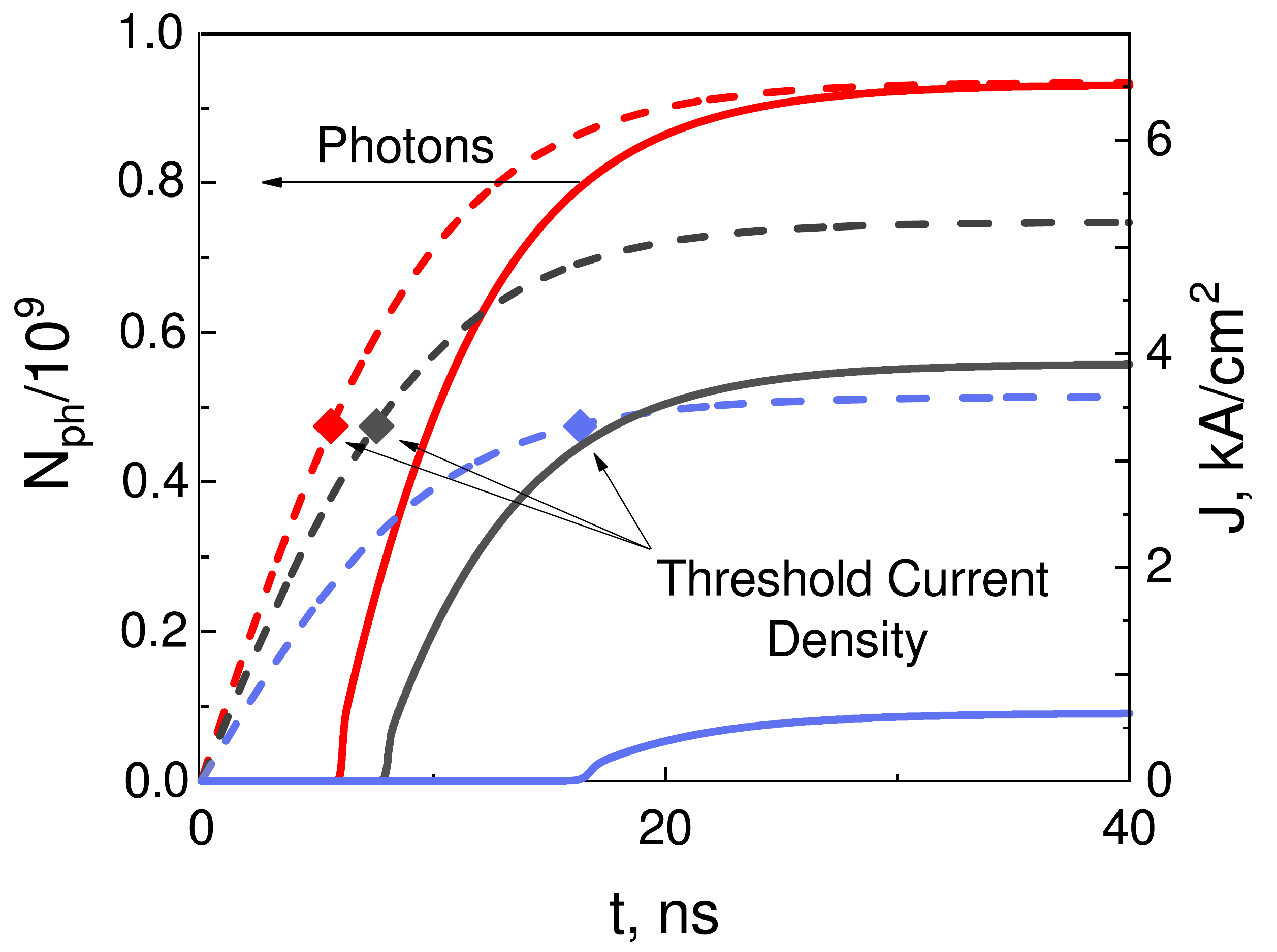}
    \caption{(color online) Pump pulses (dashed lines) used in the simulation and corresponding photon occupation numbers(solid lines) obtained through the solution of system (\ref{RateEq}). Maximum pump amplitudes are 2$J_{th}$, 1.6$J_{th}$ and 1.1$J_{th}$ (red, grey and blue lines correspondingly). Diamond marks indicate the threshold current density which is equal to 3.3 kA/cm$^2$ in the calculation.}
    \label{Pump}
\end{figure}

\begin{figure}[h]
    \centering
    \includegraphics[width=1\linewidth]{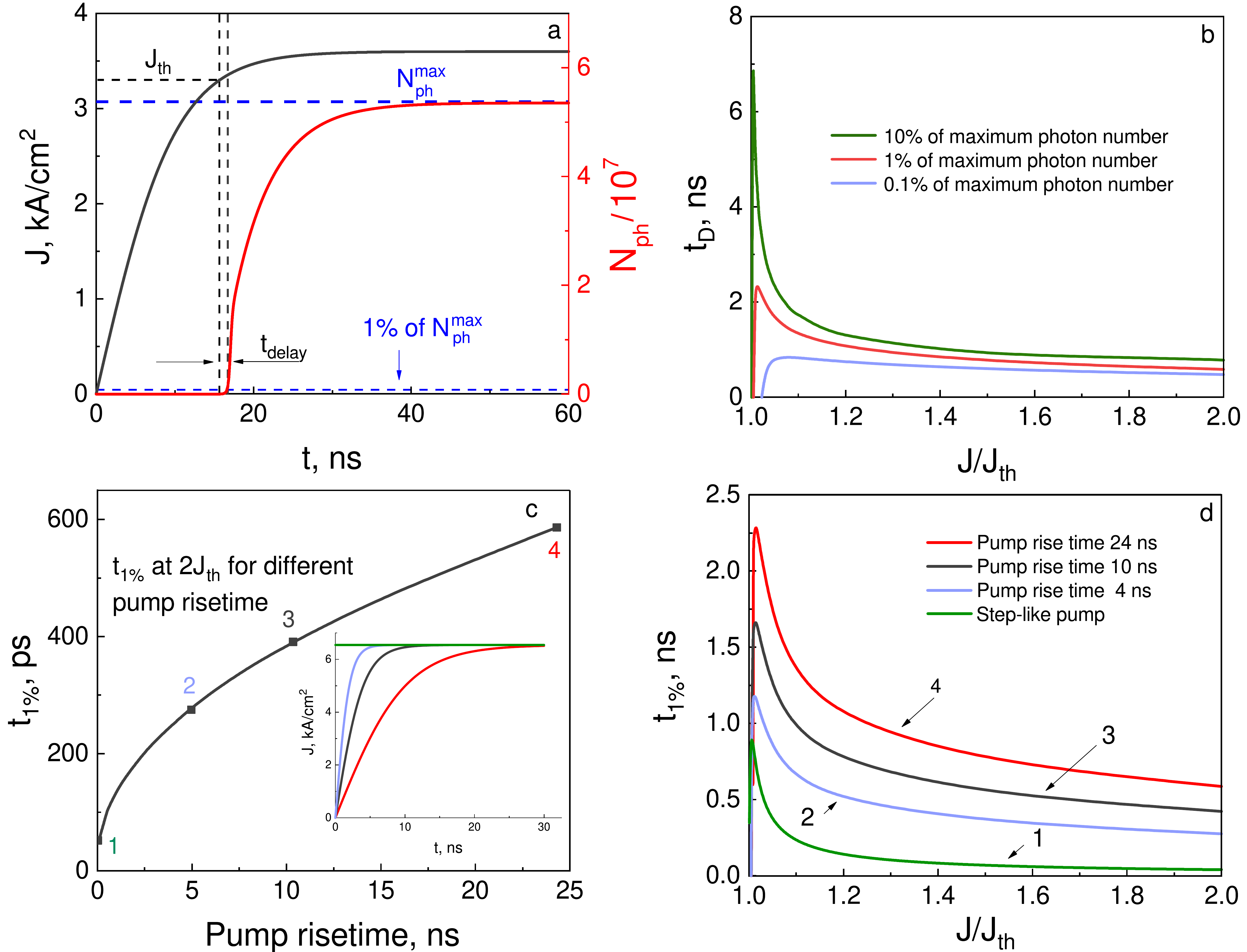}
    \caption{(color online) Results of delay time numerical simulations a) graphical definition of t$_{1\%}$ delay time. Blue lines correspond to the maximum photon occupation number and 1\% of this value. Black solid curve shows the pump pulse with $J_{max}$ = 1.1 $J_{th}$ , used in the simulation. The rise time is 24 ns. Black dashed lines indicate the threshold and delay time.The red curve is a corresponding photon occupation number growth at this time range b) delay time calculated using its different definitions: the times defined as the difference between the threshold and the moment when the photon number reaches 10{\%} (green), 1{\%}(red) and 0.1{\%} (blue) of the maximum number of photon. The pump rise time used in the modelling is 24 ns; c) t$_{1\%}$ calculated for different pump rise times at the double threshold. The inset depicts the different pump profiles used in the calculation. For the rise times marked as 1-4 we calculated t$_{1\%}$ as a function of current density shown in (d). Here, similar behavior for all pumping rise times, including zero rise time, can be seen.} 
    \label{Delay}
\end{figure}

The used model is simplified as it does not consider the carrier transport, assuming its characteristic timescales much shorter than the measured turn-on delays. Apart from that it does not describe the temporal evolution of scattering times at upper and lower levels which are current-depending below the threshold. However, all results that we claim are obtained in the region above the threshold, where the scattering times are constant due to the gain clamping. So we consider this model as applicable to simulate the delay time. 
The used parameters are taken from Ref.\cite{hamadou2009} except for the number of cascades. The latter are listed in Table~\ref{tab:table1}. The exact expression for the threshold is given in Appendix.
In the case where the pumping has a non-zero rise time, we solved this system numerically and qualitatively reproduced the results, shown in Fig.~\ref{TS5020}.
In the calculation we used the pumping with a gradual front edge given by the expression $aJ_{th}(1/(exp(-bt)+1)-c)$ where the parameter $b$ defines the rise time and the parameters $a$ and $c$ define the amplitude of the signal.
This pumping does not completely reproduce the experimental pulse shape as it does not take into account the end of the pump pulse.
Nonetheless we consider this approximation as valid, because the employed current pulses were much longer than the initial phase processes occurring in the system that were of the main interest.

The initial state is described by zero occupation numbers $N_1=N_2=N_3=N_{ph}=0$. We studied two pumping schemes.
In the first one we varied the length of the front edge of the pumping pulse while keeping the maximum amplitude constant. In this part we also numerically investigated the zero rise time pump due to the reasons mentioned in the discussion section.

In the second regime we varied the pump pulse amplitude while keeping the front rise time constant. 
The latter case is close to the experimental conditions, where the rise time of the coming pump pulse is about 20 ns, while the maximum amplitude varied in the range of 1-3 threshold values.
Fig.~\ref{Pump} depicts the shape of the electric signal used in the calculations and the corresponding calculated photon population. The rise time of this signal is 24 ns. These simulations were further used in the theoretical study of the delay time.

In Fig.~\ref{Delay} we show three types of delay values, studied in the region of pumping current densities $J/J_\mathrm{th}$ spanning the range from 1 to 2: the delay times $t_{0.1\%}$, $t_{1\%}$ and $t_{10\%}$, defined as the difference between the threshold and the moment when the photon number reaches 0.1\%, 1\% and 10\% of the maximum photon population correspondingly.
In the measurements the delay time was defined as the difference between the threshold time and the time, when the intensity reaches 1\% of the maximum photon number. This definition is further used in the paper to make the comparison with the experiment.
Apart from that we investigated the $t_{1\%}$ at the double threshold, as a function of pumping rise time (see figures \ref{Delay}c and \ref{Delay}d). It demonstrates a nonlinear growth, proving that under realistic experimental conditions, where the pump shape always has some finite steepness, one should expect the substantial increase in the turn-on delay that should be considered when designing QCL industrial applications.

\section{Discussion}
We find that the delay time dependence on the pumping current, obtained as a simulation result, is non-monotonic and has a maximum in the vicinity of the threshold pumping power.
This behavior is very different with the earlier theoretical predictions \cite{hamadou2009,hamadou2013,yong2013,yong2018,dudelev2018,agnew2016,kundu2018,choi2008,webb2014,hamadou2018,Ashok2019} and  similar to the observed dependence of the delay time on the pumping current in our experiments. We claim that this "$\Lambda$" shape of the turn-on delay dependence on pumping is due to the competition of spontaneous and stimulated emission rates in the region near the threshold (of the order of 1.01 of the threshold value). It is important to note that this competition is present for both non-zero rise time and step-like pump pulses. First, we note that for the pumping range near the threshold the signal amplitude is dominated by the spontaneous rather than stimulated emission. In addition, in the case of  non-zero rise time pumping one should keep in mind that the spontaneous emission grows linearly with the current starting from zero value, while its stimulated counterpart competes it only above the threshold (see Appendix). Therefore, spontaneous emission starts growing from the very beginning of the pumping current pulse, while the stimulated part is out of consideration until the rising edge of the pump current pulse reaches the threshold.
At weak pumping (only slightly above threshold) this leads to appearance of significant fraction of (very low) maximum output power immediately at threshold or even before the threshold is reached. This is the reason for very counter-intuitive ‘quasi-instantaneous’ response close to threshold in figures 5b,d.
The explanation given above is not applicable to the step-like pumping as the spontaneous and stimulated emission grow simultaneously in this case. However, for pumping amplitudes very close to the threshold the rate of spontaneous emission is comparable or even higher than that of its stimulated counterpart, which is low at low photon occupation numbers.
One can see that from the numerical simulation results at even lower pumping than discussed above, of the order of 1.001 threshold value at given parameters, as shown with green curve in figure 5d.
With these results we conclude that in order to quantitatively reproduce the experimental measurements, the exact values for the parameters used in the rate equations solution should be carefully measured.

\section{Conclusion}
We presented a study of the turn-on delay of lasing in mid-infrared QCLs based on different materials. In QCLs of both types the delay times were found to be orders of magnitude longer than the values predicted by the existing models considering step-like excitation.Also, the observed dependence of the delay time on the pumping current was found to have unexpected "$\Lambda$" shape, in contrast to the theoretically predicted montonous decrease with pumping. We proposed a theoretical model of the QCL response to pump pulses with non-zero rise time, which gave a fair agreement with the observed dependence of the turn-on delay on the pump intensity. The dependence of the turn-on delay vs pump rise time shows that there is an optimal region of the front rise-time, reachable by the modern current sources, that yeilds the turn-on delay below 400ps. We believe that further investigation of structural parameters responsible for the key difference between InAs- and InP-based QCL will also help to optimize the QCL design and decrease the turn-on delay.

\appendices
\section{Modelling details}
Here we present the details of the numerical modelling, that are not shown in the main manuscript. In the calculation, the threshold current was defined as follows (see the ref. \cite{hamadou2009} for the details):
\begin{equation}
J_{th} = \frac{1}{\tau_{3}\left( 1-\frac{\tau_{21}}{\tau_{32}}-\frac{\tau_{21}}{\tau_{sp}}\right) }\frac{L\epsilon_0n_{eff}\lambda(2\gamma_{32})}{4\pi ez_{32}^2}\frac{(\alpha_w+\alpha_m)}{\Gamma}
\end{equation}
where $\alpha_w$ and $\alpha_m$ are waveguide and mirror losses correspondingly, $ez_{32}$ is the dipole matrix element of the transition between the states 3 and 2, $2\gamma_{32}$ is the full width at half maximum of the electroluminescence spectrum, other parameters are listed in Table 1 of the main manuscript.
The graphs \ref{S1} and \ref{S2} below show the temporal evolution of the occupation numbers of all the levels and photonic mode and their dependence on the spontaneous emission coefficient $\beta$.

\begin{figure*}[h]
\centering
\includegraphics[width=0.8\linewidth]{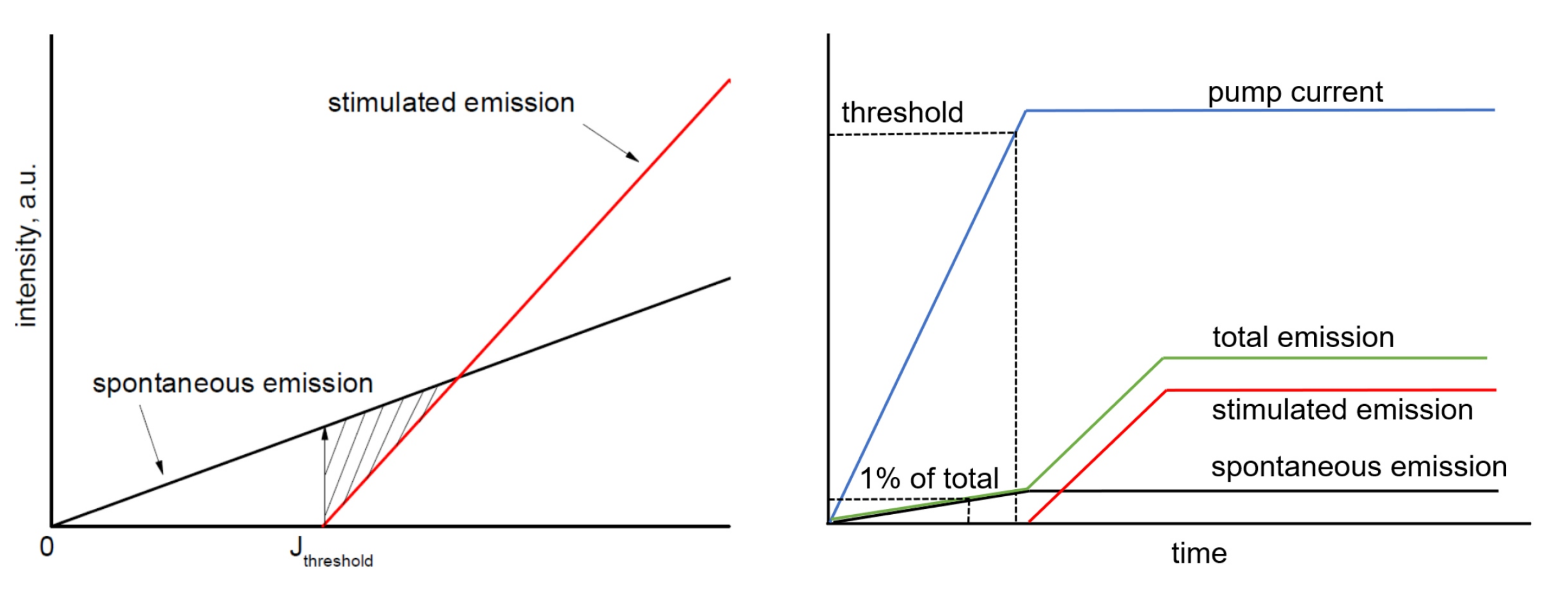}
\caption{The schematic illustration of the interplay between spontaneous and stimulated emission. Left panel shows the region, where the spontaneous photons exceed the stimulated ones. Right panel qualitatively shows the overall contribution of spontaneous and stimulated emission as a function of time.}
\label{Sup4}
\end{figure*}

\begin{figure}[h]
\centering
\includegraphics[width=0.9\linewidth]{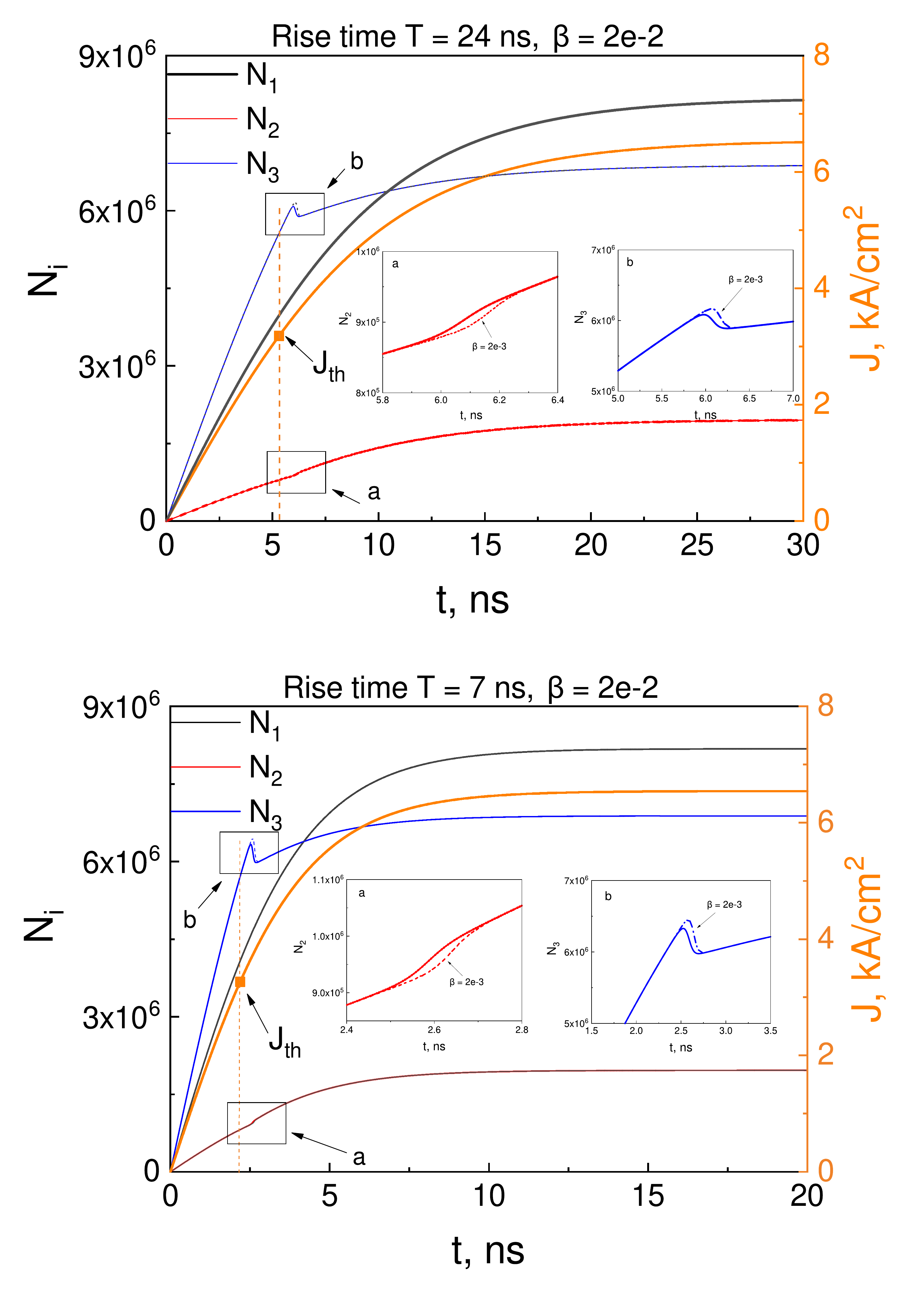}
\caption{The temporal evolution of the levels $N_1, N_2$ and $N_3$, for the pumping rise time T = 24 ns (upper panel) and 7ns (lower panel). In the main figure $\beta$ = $2*10^{-2}$, while the insets enlarge the areas, where the difference between $\beta$ = $2*10^{-2}$ and $\beta$ = $2*10^{-3}$ can be observed. Orange curve shows the pump pulse current density. Orange dashed line indicates the threshold time and current density The maximum current amplitude is $J=2J_{th}$}
\label{S1}
\end{figure}
\begin{figure}[h]
\centering
\includegraphics[width=0.8\linewidth]{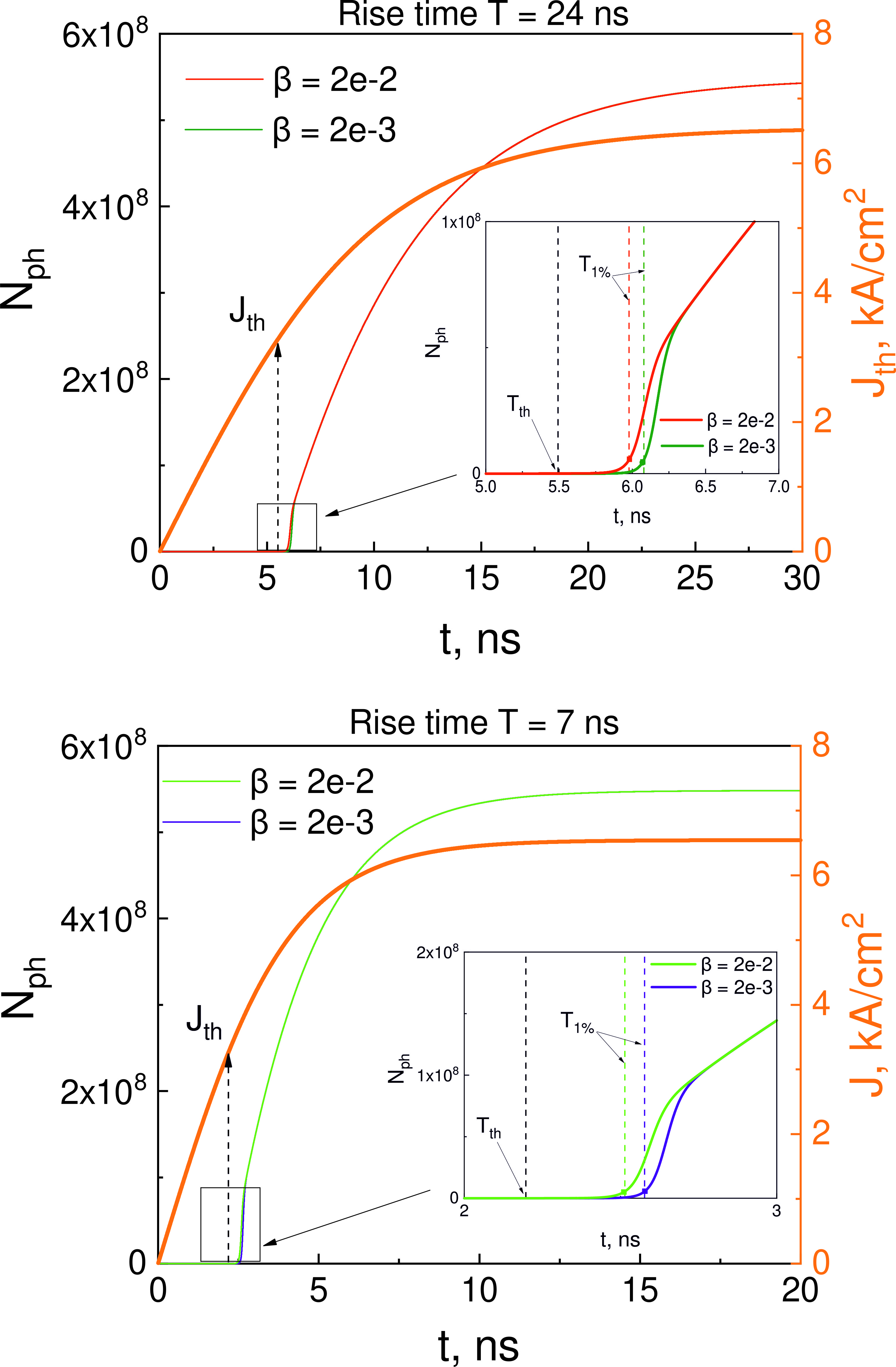}
\caption{The temporal evolution of the photonic mode $N_{ph}$, for the pumping rise time T = 24 ns (upper panel and 7ns (lower panel). The insets enlarge the areas, where the difference between $\beta$ = $2*10^{-2}$ and $\beta$ = $2*10^{-3}$ can be clearly observed. The maximum current amplitude is $J=2J_{th}$. Orange curve shows the pump pulse current density. Dashed lines indicate the threshold time (black) and the time, when the photon occupation number reaches 1\% of its maximum.(red-green (a) and violet-green (b))}
\label{S2}
\end{figure}

It is shown that the lower spontaneous emission is, the longer is the delay. However, to reveal the role of spontaneous emission in full one should investigate the near-threshold interplay of spontaneous and stimulated emission.

In Fig.~\ref{Sup4} we present the schematic illustration of this idea. In the pumping range near the threshold the maximum signal amplitude largely consists of the spontaneous rather than stimulated emission. Spontaneous emission starts from the very beginning of the pumping current pulse while the stimulated part doesn’t start before the threshold is reached by the rising edge of the pump current pulse. At low pumping (only slightly above threshold) that leads to appearance of significant fraction of (very low) maximum output power immediately at threshold or even before the threshold is reached. The discussed 1$\%$ concentration then mostly contains the spontaneous photons. The exact behavior should be studied taking into account all parameters used in the calculation.

\section*{Acknowledgment}
Part of this work was funded by French Program  “Investment for the Future” (Equipex EXTRA, ANR-11-EQPX-0016). The authors from Ioffe Institute acknowledge the support from Russian Science Foundation (project 21-72-30020)
E.C. thanks Dr. A. Nalitov for valuable discussions

\ifCLASSOPTIONcaptionsoff
  \newpage
\fi



%

\bibliographystyle{IEEEtran}
\bibliography{IEEEabrv,bare_jrnl}



%








\end{document}